\documentclass[aps,prl,twocolumn,showpacs,superscriptaddress]{revtex4}  
\usepackage{graphicx}  
\usepackage{dcolumn}   
\usepackage{bm}        
\usepackage{amssymb}   
\usepackage{amsfonts}  
\usepackage{hyperref}

\input babarsym

\newcommand{\BABARPubYear}    {11}
\newcommand{\BABARPubNumber}  {011}
\newcommand{\SLACPubNumber} {14659}

\begin{document}

\leftline{\babar-PUB-\BABARPubYear/\BABARPubNumber }
\rightline{SLAC-PUB-\SLACPubNumber }

\title{Search for $\overline{\bm{B}} \to \bm{\Lambda_c^+ X \ell^-} \overline{\bm{\nu}}_{\bm{\ell}}$ Decays in Events With a Fully Reconstructed $\bm{B}$ Meson}

%
\author{J.~P.~Lees}
\author{V.~Poireau}
\author{V.~Tisserand}
\affiliation{Laboratoire d'Annecy-le-Vieux de Physique des Particules (LAPP), Universit\'e de Savoie, CNRS/IN2P3,  F-74941 Annecy-Le-Vieux, France}
\author{J.~Garra~Tico}
\author{E.~Grauges}
\affiliation{Universitat de Barcelona, Facultat de Fisica, Departament ECM, E-08028 Barcelona, Spain }
\author{M.~Martinelli$^{ab}$}
\author{D.~A.~Milanes$^{a}$}
\author{A.~Palano$^{ab}$ }
\author{M.~Pappagallo$^{ab}$ }
\affiliation{INFN Sezione di Bari$^{a}$; Dipartimento di Fisica, Universit\`a di Bari$^{b}$, I-70126 Bari, Italy }
\author{G.~Eigen}
\author{B.~Stugu}
\author{L.~Sun}
\affiliation{University of Bergen, Institute of Physics, N-5007 Bergen, Norway }
% ...
\collaboration{The \babar\ Collaboration}
\noaffiliation

\date{\today}

\begin{abstract}
We present a search for semileptonic $B$ decays to the charmed baryon $\Lambda_c^+$ based on 
420 fb$^{-1}$ of data collected at the $\Upsilon(4S)$ resonance with the \babar\
detector at the \pep2\ $e^+e^-$ storage rings.
By fully reconstructing the recoiling $B$ in a hadronic decay mode,
we reduce non-$B$ backgrounds and determine the flavor of the signal $B$.
We statistically correct the flavor for the effect of the $B^0$ mixing.
We obtain
a 90\% confidence level upper limit of
${\cal B}(\Bbar \to \Lambda_c^+ X \ell^- \overline{\nu}_{\ell})
  /{\cal B}(\Bbar \to  \Lambda_c^+ X) < 3.5 \% $.
\end{abstract}

\maketitle 

\begin{figure}[hbtp]
\begin{center}
\includegraphics[width=0.23\textwidth]{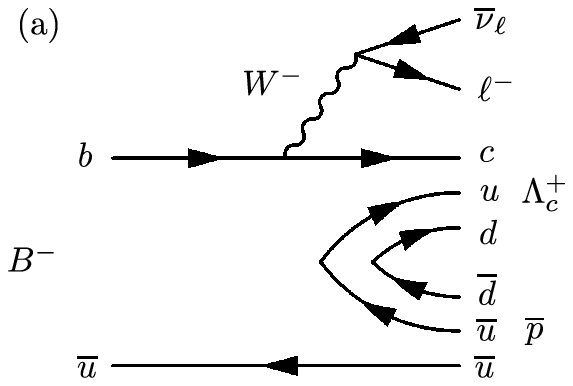}
\hfill
\includegraphics[width=0.23\textwidth]{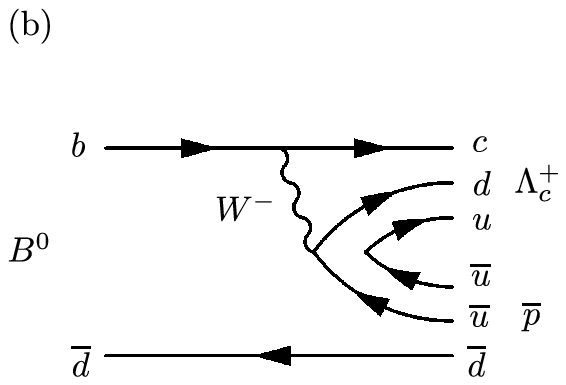}
\end{center}
\caption{Feynman diagrams for $B$ decays into a charmed baryon through external $W$ emission (a) 
and internal $W$ emission (b).}
\label{fig_diagram}
\end{figure}

Decays of $B$ mesons to charmed baryons are not as well understood as those to charmed mesons.  
In particular, there is limited knowledge, both theoretical and experimental, about semileptonic 
$\Bbar$ decays to the $\Lambda_c^+$ charmed baryon~\cite{ell}.
If $\overline{B}$ decays to charmed baryons are dominated 
by external $W$ emission (Fig.~\ref{fig_diagram}a),
as is the case for $\overline{B}$ decays to charmed
mesons~\cite{introCites,introCites2},
and final-state hadronic interactions are small,
the semileptonic fraction of these decays should be roughly the same:
\begin{equation}	
\frac{{\mathcal B}(\Bbar \to \Lambda_c^+ X \ell^- \overline{\nu}_{\ell})}
     {{\mathcal B}(\Bbar \to  \Lambda_c^+ X^\prime)}
\sim
\frac{{\mathcal B}(\Bbar \to D X^{\prime\prime} \ell^- \overline{\nu}_{\ell})}
     {{\mathcal B}(\Bbar \to D X^{\prime\prime\prime})}
\end{equation}
where $\ell = e$ or $\mu$,
and $D$ is understood to be $D^{(*)0}$ or $D^{(*)+}$.
The semileptonic fraction of $B$ decays to charmed mesons is 
currently measured to be $11.1\pm0.8$\%~\cite{Nakamura:2010zzi}.
A significantly smaller semileptonic ratio 
for $B$ decays to charmed baryons would be evidence for
a sizable internal $W$ emission amplitude
in baryonic $B$ decay (Fig.~\ref{fig_diagram}b)
or a large contribution of final state interactions.

About 90\% of the measured inclusive semileptonic $\Bbar \to X_c \ell^- \overline{\nu}_{\ell}$ branching 
fraction into charmed final states can be accounted for by summing the branching fractions 
from exclusive $\Bbar \rightarrow D^{(*)}(\pi) \ell^- \overline{\nu}_{\ell}$ decays~\cite{Aubert:2007qw}.
Semileptonic $B$ decays to  charmed baryons could account for some of the remaining difference.

A previous search for semileptonic $B$ decays into charmed baryons by the CLEO 
collaboration~\cite{Bonvicini:1997fn} resulted in an upper limit on the ratio
${\cal B}(\Bbar \to \Lambda_c^+ X e^-\overline{\nu}_e)/ {\cal B}
(B/\Bbar \to  \Lambda_c^+ X)<5\%$ at the 90\% confidence level.  
By using ${\cal B}(B/\Bbar \to  \Lambda_c^+ X) = 0.045 \pm 0.004 \pm 0.012$ 
and ${\cal B}(\Bbar \to  \overline{\Lambda}_c^- X)/{\cal B}(\Bbar \to  \Lambda_c^+ X) = 0.19 \pm 0.13 \pm 0.04$~\cite{Nakamura:2010zzi},
and assuming lepton universality,
this result implies a semileptonic fraction limit 
${\cal B}(\Bbar \to \Lambda_c^+ X \ell^- \overline{\nu}_{\ell})/{\cal B}(\Bbar \to  \Lambda_c^+ X) < 6\%$ 
at 90\% confidence level.

There are two caveats to the CLEO measurement.
First, the electron candidate is required to have a momentum greater than 0.6~GeV/$c$,
which reduces background due to fake and secondary electrons,
but may also reduce signal efficiency.
Second, because the CLEO measurement was unable to
constrain the flavor of the $\overline{B}$ meson,
the quoted fraction suffers from large systematic uncertainties
due to the uncorrelated
$\overline{B}\to \overline{\Lambda}_c^-X$ background.
We address these two points by reconstructing a $B$ meson in
a hadronic mode and look for the signal in its recoil.
The resulting sample has less background, which
allows us to lower the lepton momentum cutoff,
and the flavor of the hadronic $B$ meson determines the
flavor of the signal $\overline{B}$, up to mixing effects.
By normalizing to the correlated $\Bbar \to \Lambda_c^+X$ decay mode,
many systematic uncertainties cancel.

In this paper, we present a search for semileptonic
$\Bbar$ decays to $\Lambda_c^+$ using data collected 
with the \babar\ detector at the 
\pep2\ asymmetric-energy $e^+e^-$ storage rings at SLAC.  The data consist of a total 
of 420~fb$^{-1}$ recorded at the $\Upsilon(4S)$ resonance between 1999 and 2008,
corresponding to approximately 460 million \BB\ pairs.
The \babar\ detector is described in detail elsewhere~\cite{Aubert:2001tu}.
Charged particle trajectories
are measured by a five-layer double-sided silicon vertex tracker (SVT)
and a 40-layer drift chamber (DCH), 
both operating in a 1.5 T magnetic field.
Charged particle identification is provided by the 
specific ionization energy loss (d$E$/d$x$) in the
tracking devices and by an internally reflecting ring-imaging Cherenkov detector (DIRC). 
Photons are detected by a CsI(Tl) electromagnetic calorimeter (EMC).
Muons are identified by the instrumented magnetic-flux return (IFR).
A detailed Geant4-based Monte Carlo (MC) simulation~\cite{Agostinelli:2002hh} of \BB\ and
continuum events (light quarks and $\tau$ pairs) is used to study the detector response,
its acceptance,
and to test the analysis techniques.

We search for semileptonic $\Bbar \to \Lambda_c^+ X \ell^-\overline{\nu}_{\ell}$ 
decays
with $\ell=e$ or $\mu$
in events pre-selected to contain a candidate $B$ reconstructed
in a fully hadronic decay mode ($B_{\rm tag}$), as described later in the text.  We select signal candidates
in these events by looking for candidate leptons and fully reconstructed $\Lambda_c^+$ decays.
We then refine our selection of $B_\mathrm{tag}$, and make a final signal extraction based on
the selected $B_\mathrm{tag}$ and $\Lambda_c^+$ kinematic properties.  
We also select candidate $\Bbar \to \Lambda_c^+ X$ events, starting with the same sample 
and using similar techniques and selections,
but without requiring an identified lepton candidate.

Selection criteria are optimized using MC simulation of signal and background processes.
Because little is known about $\Bbar \to \Lambda_c^+ X \ell^-\overline{\nu}_{\ell}$ decays,
we use a signal model which can be tuned to cover a large
range of possible kinematics of the final-state particles.
In this model,
the $\Bbar$ decays semileptonically into an intermediate massive particle 
$Y$, $\Bbar \to Y \ell^-\overline{\nu}_{\ell}$, with a kinematic distribution according 
to phase space~\cite{Sjostrand:1993yb}.
The $Y$ subsequently decays into a $\Lambda_c^+$, an anti-nucleon (anti-proton or anti-neutron),
and $n_1$ ($n_2$) charged (neutral) pions, again assuming phase space distributions.
The free parameters in the model (the mass $m_Y$ and width $\Gamma_Y$ of the pseudo-particle $Y$,  
and $n_1$ and $n_2$) are tuned to reproduce the 
lepton and charmed hadron momentum spectra predicted by the
$\Bbar \to D^{(*)}\pi \ell \overline{\nu}_{\ell}$ model of Goity and Roberts~\cite{Goity:1994xn},
after accounting for the phase space limits implied by the large baryon masses.  We choose $m_Y=4.5$
GeV/$c^2$, $\Gamma_Y=0.2$ GeV/$c^2$, and $n_1 +n_2 \leq 6$.

We reconstruct $B_{\rm tag}$ decays of the type $B \rightarrow \Dbar Y'$,
where $Y'$ represents a collection of hadrons with a total charge of $\pm 1$, composed 
of $n_1'\pi^{\pm}+n_2' K^{\pm}+n_3' K^0_S+n_4'\pi^0$,
   where $n_1'+n_2' \leq  5$, $n_3' \leq 2$,  and $n_4' \leq 2$.
$K^0_S$ candidates are reconstructed in the $\pi^+\pi^-$ decay mode,
$\pi^0$ candidates in the $\gamma\gamma$ mode.
Using $\overline{D}^0$ $(D^-)$ and $\overline{D}^{*0}$ $(D^{*-})$ as seeds for
$B^+$ $(\Bz)$ decays, we reconstruct about 
1000 complete $B$ decay chains~\cite{Aubert:2003zw}.

The kinematic consistency of a $B_\mathrm{tag}$ candidate with a $B$ meson decay 
is evaluated using two variables: the beam-energy
substituted mass $m_{ES} \equiv \sqrt{s/4-|p^*_B|^2}$, and the energy difference 
$\Delta E \equiv E^*_B -\sqrt{s}/2$. Here $\sqrt{s}$ is the
 total center of mass (CM) energy, and $p^*_B$ and $E^*_B$ denote the momentum and energy of 
 the $B_\mathrm{tag}$ candidate in the CM frame. For correctly identified 
 $B_\mathrm{tag}$ decays, the $m_{ES}$ distribution peaks at the $B$ meson mass,
with a resolution of about 2.5 MeV/$c^2$
averaged over the decay modes,
while $\Delta E$ is consistent
with zero, with a resolution of about 18 MeV.
We select $B_\mathrm{tag}$ candidates in the signal region
defined as 5.27~GeV/$c^2$ $< m_{ES} <$ 5.29~GeV/$c^2$,
with a $\Delta E$ within $4 \sigma$ of zero.
This selection has an estimated efficiency of 0.2\% to 0.3\% per $B$ meson.

We identify electron and muon candidates by combining the information
on the measured momentum and energy loss in the SVT and DCH,
the angle of Cherenkov radiation in the DIRC,
and the energy deposition and shower shape in the EMC.
For sufficiently hard muons, the information from the IFR is also used.
We correct for bremsstrahlung of electrons by combining the
four-momentum of the electron with those of detected photons
which are emitted close to the electron direction.
We require lepton candidates to have a
momentum in the CM frame $p^*_{\ell} > 0.35$ GeV/$c$
and a point of closest approach to the 
collision axis of less than 0.1 cm.  The 
$p^*_{\ell}$ selection value is motivated by the large mass of the 
$\Lambda_c^+$ and the assumption of another baryon in the decay due to baryon number
conservation, which greatly restricts the kinetic energy available to the leptons.
We identify photon conversions and $\pi^0$ Dalitz decays
using a dedicated algorithm based on the vertex and kinematic properties of 
two opposite charge tracks, and eliminate electron candidates coming from these.

Candidate $\Lambda_c^+$ baryons
are reconstructed in the $p K^-\pi^+$, $p K^0_S$, $p K^0_S \pi^+
\pi^-$, $\Lambda \pi^+$,
and $\Lambda \pi^+ \pi^+ \pi^-$ modes.
$\Lambda$ candidates are reconstructed in the $p\pi^-$ decay mode.
Only $\Lambda_c$ candidates with opposite charge of the lepton candidate are considered.
Charged daughters of the $\Lambda_c^+$
candidate are fit to a vertex tree~\cite{Hulsbergen:2005pu}, with
$K^0_S$ and $\Lambda$ masses constrained to their known values~\cite{Nakamura:2010zzi}, and
the $\Lambda_c^+$ origin
constrained to the known average luminous position
of the beams
within its measured size and uncertainties.
In events with multiple $\Lambda_c^+$ and/or $\ell$ candidates, the candidates are fit
to a common vertex, and the $\Lambda_c^+\ell^-$ pair with the highest
vertex fit probability is selected.

We refine the selection of $B_\mathrm{tag}$ candidates by first removing those whose
daughter particles are based on tracks already used to reconstruct the
signal-side $\Lambda_c^+$ or lepton and those charged $B_\mathrm{tag}$ candidates whose
flavor is
opposite that of the signal $\overline{B}$ candidate.
We account for mixing effects by weighting $\Bz$ and $\Bzb$ tags
according to the $\Lambda_c$ charge, as described in Ref.~\cite{Aubert:2004td}.
In events with multiple $B_{\rm tag}$ candidates,
we select the one reconstructed in the highest purity mode,
where the purity is estimated for each $B_{\rm tag}$ decay chain
using MC simulation as the ratio of signal over background events.  
When multiple candidates in the same event have the same $B_{\rm tag}$ mode,
we select the one with the smallest $|\Delta E|$ value.

We reconstruct the CM missing momentum $\vec{p}_\mathrm{miss}$ by noting that
$\vec{p}_\mathrm{miss} + \vec{p}_\mathrm{vis} = \vec{0}$
in the CM frame, where the visible momentum $\vec{p}_\mathrm{vis}$
is computed by summing the momentum vectors
of the $B_\mathrm{tag}$, the $\Lambda_c$ and $\ell$ candidates, plus any additional well measured
charged track or neutral cluster boosted to the CM frame.
We require $|\vec{p}_\mathrm{miss}| > 0.2$ GeV$/c$
to remove background from hadronic $\Bbar \to \Lambda_c^+ X$ decays in
which all the particles in the $X$ system have been reconstructed and one
hadron is misidentified as a lepton.  We compute the total observed charge of
the selected events by adding the charges of all particles used in the
$\vec{p}_\mathrm{miss}$ calculation, and require this
to be zero. This reduces the background in the $B_{\rm tag}$
reconstruction due to missing particles.

Backgrounds are divided according to whether they contain
a correctly reconstructed $\Lambda_c^+$ candidate.
Those which contain such a candidate are called ``peaking background,''
while those that do not are called ``combinatorial background.''
The predictions from MC simulation of generic \BB\ and continuum events
show that the peaking background
arises mainly from hadronic $\Bbar \to \Lambda_c^+ X$ decays, where
the $\Lambda_c^+$ is correctly reconstructed, and the lepton candidate is
an electron from gamma conversions or $\pi^0$ Dalitz decays, or a hadron
misidentified as a muon; we estimate
$3.6\pm 0.7_{\rm stat.}\pm 0.7_{\rm syst.}$ and
$15.3\pm 1.5_{\rm stat.}\pm 1.4_{\rm syst.}$ peaking background events for the
electron and muon samples, respectively.
The relatively large peaking background rate for the muon channel
is due primarily to the low lepton momentum cut.

\begin{figure}[!ht]
\centering
\includegraphics[width=0.5\textwidth]{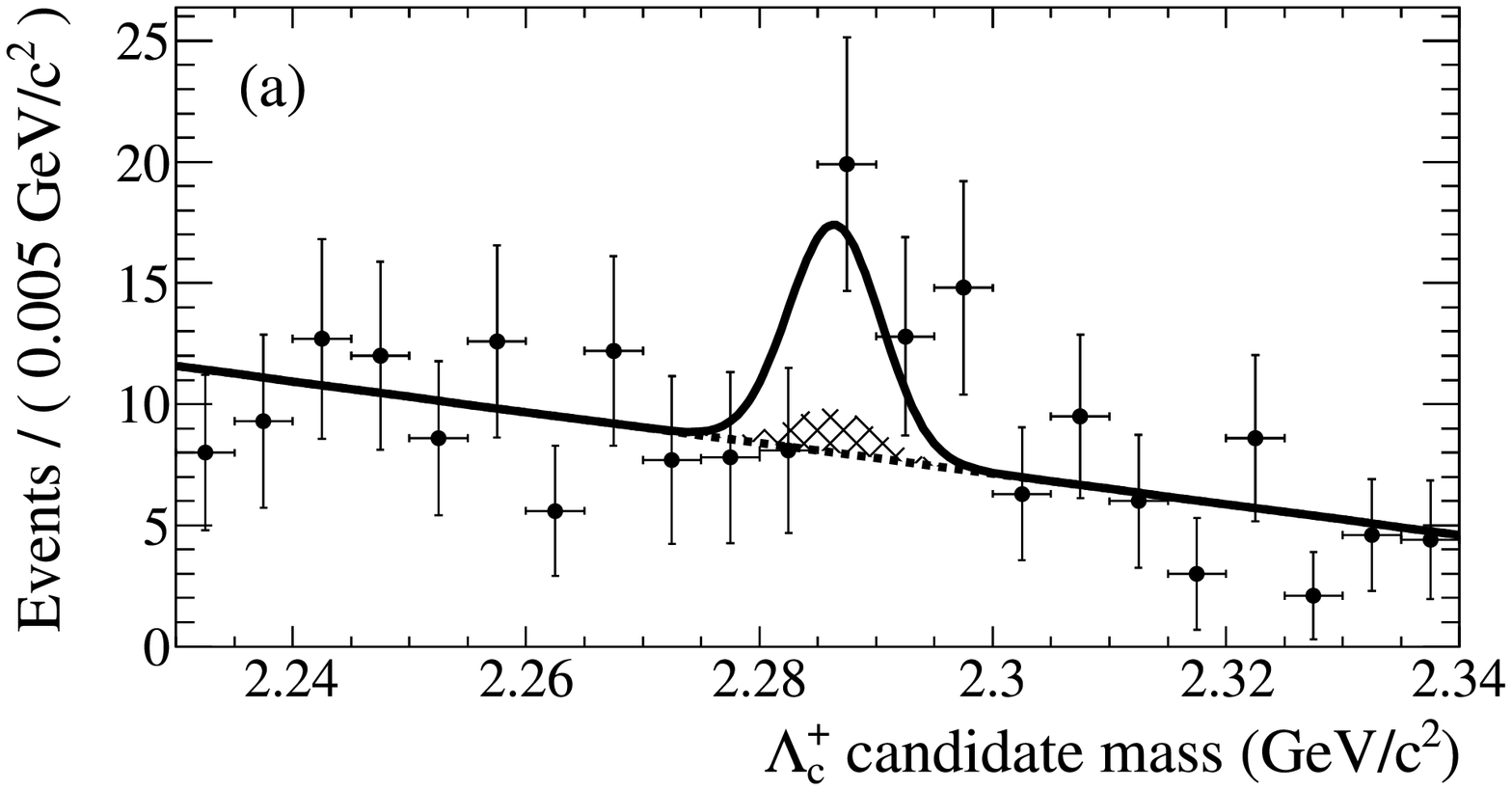}
\includegraphics[width=0.5\textwidth]{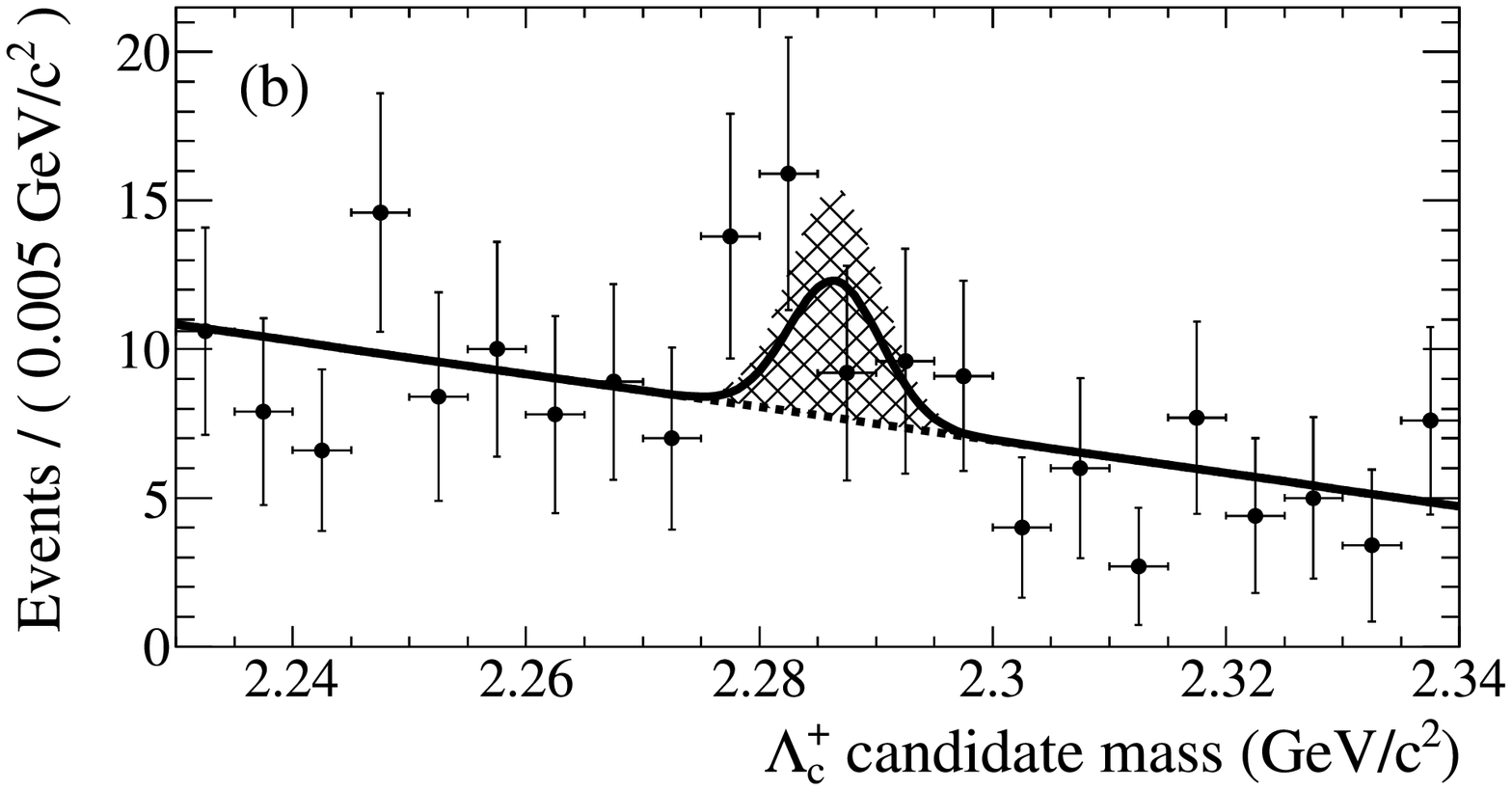}
\caption{Fit to the $\Lambda_c^+$ candidate mass distribution for
$\Bbar \to \Lambda_c^+ X e^- \overline{\nu}_{e}$ (a) and
$\Bbar \to \Lambda_c^+ X \mu^- \overline{\nu}_{\mu}$ (b).
The data are shown as points with error bars, the overall
fit as a solid line, and the peaking background contribution as a cross-hatched area.
The combinatorial \BB\ and continuum background is shown
as the area below the dotted line.}
\label{fig:Fit1}
\end{figure}

We determine the $\Bbar$ semileptonic signal yield with a simultaneous
unbinned maximum likelihood fit to the distribution of the $\Lambda_c^+$ invariant mass
on both the electron and muon samples.
The $\Lambda_c^+$ invariant-mass distribution is described by the sum of three
probability density functions (PDFs) representing signal, peaking background, and
combinatorial background.  
The functional forms of the PDFs are chosen based on simulation studies.
The signal and peaking background contributions are modeled
as Gaussian functions whose mean and width are fixed to the values
obtained from a fit to the the $\Lambda_c^+$ candidate mass spectrum
in the $\Bbar \to\Lambda_c^+X$ data sample described below.
The number of peaking background
events is fixed to the prediction from MC simulations.
The combinatorial \BB\ and continuum backgrounds
are modeled as a first-order polynomial, whose parameters are constrained by a 
fit to the $\Lambda_c^+$ invariant mass
sidebands, defined as the mass ranges from $2.23-2.26$ and $2.31-2.34$ GeV/$c^2$.
The fit to the $\Lambda_c^+$ invariant mass
is shown in Fig.~\ref{fig:Fit1},
projected separately for the electron and muon samples.  
The corresponding yields are shown in Table~\ref{tab:correlated}.

In order to reduce systematic uncertainties due to $B_\mathrm{tag}$ 
and $\Lambda_c^+$ reconstruction, the
$\Bbar \rightarrow \Lambda_c^+ X \ell^- \overline{\nu}_{\ell}$ branching
fraction is measured relative to the inclusive
${\cal B} (\Bbar \rightarrow \Lambda_c^+ X )$ branching
fraction.  To determine the inclusive yield, we start with
the same $B_\mathrm{tag}$ sample used for the semileptonic selection.  We reconstruct
$\Lambda_c^+$ candidates as in the semileptonic case, choosing
the candidate with the highest vertex probability in case of multiple candidates.
We exclude $B_{\rm tag}$ candidates
with daughter particles in common with the $\Lambda_c^+$ candidate
and resolve multiple $B_{\rm tag}$ candidates as in the semileptonic case.

\begin{figure}[!ht]
\centering
\includegraphics[width=0.5\textwidth]{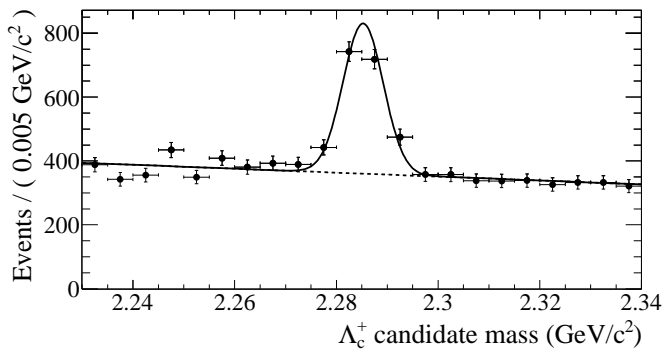}
\caption{Fit to the $\Lambda_c^+$ candidate mass distribution for
$\Bbar \to \Lambda_c^+ X$.  The data are shown as points with error bars,
the overall fit as a solid line, and the
combinatorial \BB\ and continuum background as a dashed line.}
\label{fig:Fit2}
\end{figure}

We determine the $\Bbar \to \Lambda_c^+ X$ signal yield with an
unbinned maximum likelihood fit to the $\Lambda_c^+$ invariant mass.
The fit function consists of the sum of two
PDFs representing signal and combinatorial background, described by
a single Gaussian and a first order polynomial, respectively.
All parameters of the signal Gaussian
are left free in the fit.  We obtain a $\Lambda_c^+$ mass value of
$2.2853 \pm 0.0003$ GeV/$c^2$, consistent with the current world average~\cite{Nakamura:2010zzi},
and a resolution of $4.0\pm0.3$ MeV/$c^2$, consistent with expectations from MC simulations.
The $\Lambda_c^+$ invariant mass distribution on the inclusive sample
and the results of the fit are shown in Fig.~\ref{fig:Fit2}.

We determine the relative branching fraction
${\cal B} (\Bbar \to \Lambda_c^+ X\ell^- \overline{\nu}_{\ell})/
{\cal B} (\Bbar \to \Lambda_c^+ X)$ as the ratio of the measured signal yields,
after correcting for the ratio of the reconstruction efficiencies:
\begin{equation}
\frac{{\cal B}(\Bbar \to\Lambda_c^+X\ell^- \overline{\nu}_{\ell})}{{\cal B}( \Bbar\to\Lambda_c^+ X)} =
\left(\frac{N_{\rm s}}{N_{\rm i}}\right)
\left(\frac{\epsilon_{\rm i}}
{\epsilon_{\rm s}}\right).
\end{equation}
Here, $N_{\rm s}$  ($N_{\rm i}$) is the number of
$\Bbar \to \Lambda_c^+ X\ell^-\overline{\nu}_\ell$ ($\Bbar \to \Lambda_c^+ X$)
events reported in Table~\ref{tab:correlated}
together with the corresponding reconstruction
efficiencies $\epsilon_{\rm s}$ ($\epsilon_{\rm i}$);
the latter include the $B_{\rm tag}$ efficiencies,
which are estimated with MC simulation.

Many systematic uncertainties approximately cancel in this ratio,
such as those due to the $\Lambda_c^+$ and $B_{\rm tag}$ reconstruction efficiencies and the
$\Lambda_c^+$ decay branching fractions.
We categorize the remaining systematic uncertainties into those
which directly affect the signal yield, and those
which affect only the efficiency.
The systematic uncertainties that have been considered are described below and
summarized in Table~\ref{tab:r24-systematic}.

Systematic uncertainties in the signal yield are dominated by the peaking background yield.
We estimate this by propagating the uncertainty in the $\Bbar \to\Lambda_c^+ X$ branching fraction,
and the Poisson error from the MC simulation.
We add in quadrature the effect of varying the
probability for a pion to be misidentified as
an electron or as a muon by 15\%,
where the range is estimated using data control samples~\cite{Aubert:2003zw}.
Systematic uncertainties due to
background electrons from photon conversions
and $\pi^0$ Dalitz decays are negligible.

To account for a possible bias due to the fit technique,
we prepare ensembles of MC experiments, 
in which events are generated according to
the PDF shapes determined from data.
We vary the signal to background ratio
and fit for the signal as in the full analysis.  The average difference between
the fitted value of the yield and the true value is taken as a systematic uncertainty,
labeled ``Fit bias'' in Table~\ref{tab:r24-systematic}.

\begin{table}[!htb]
\caption{Signal yields and reconstruction efficiencies for the
$\Bbar \to \Lambda_c^+ X \ell^- \overline{\nu}_{\ell}$,
$\Bbar \to \Lambda_c^+  X$,
and
$B/\Bbar \to \Lambda_c^+  X$
decays with the corresponding statistical uncertainties.}
\label{tab:correlated}
\centering
\begin{tabular}{lcc}
\hline
\hline \\[-0.98em]
Decay mode & $N_{\rm data}$ & $\epsilon$ $(\times 10^{-5})$  \\
\hline \\[-0.98em]
$\Bbar \rightarrow \Lambda_c^+ X e^- \overline{\nu}_{e}$
& $15.0 \pm 6.8$
& $1.98\pm 0.17$ \\
$\Bbar \rightarrow \Lambda_c^+ X \mu^- \overline{\nu}_{\mu}$
& $-6.2 \pm 6.3$
& $1.04\pm 0.12$ \\
$\Bbar \rightarrow \Lambda_c^+ X$
& $934 \pm 55$
& $3.09\pm 0.11$ \\
$B/\Bbar \rightarrow \Lambda_c^+ X$
& $1386 \pm 66$
& $3.21\pm 0.12$ \\
\hline
\hline
\end{tabular}
\end{table}

Systematic uncertainties on the reconstruction efficiency ratio
are dominated by the uncertainty in the signal model.
This is estimated by comparing our nominal signal model with a pure phase space model,
where the $\Bbar \to\Lambda_c^+X\ell^- \overline{\nu}_{\ell}$ decay occurs in one step,
taking the full difference in the signal efficiency estimate compared to our
nominal signal model as the systematic uncertainty.
The larger systematic uncertainty for the muon channel
is due to the low muon identification efficiency for the soft leptons.
The uncertainty in the reconstruction efficiency
due to the limited statistics of the MC simulation
is added as a systematic uncertainty
by weighting the events to the data size.
The peaking background in the inclusive mode
due to $c\overline{c}$ is estimated using the prediction from our MC simulation
and is found to be compatible with the statistical uncertainty of the sample,
which we take as a systematic uncertainty.
We estimate the systematic uncertainty on the signal efficiency due to particle identification
by varying the electron (muon) identification efficiency by 2\% (3\%),
based on studies using data control samples~\cite{Aubert:2003zw}.
Since the order for selecting the best candidate is different
between the semileptonic and inclusive samples,
the uncertainties on the ratio of the $B_{\rm tag}$ and $\Lambda_c^+$ efficiencies do not exactly cancel.
We evaluate the corresponding systematic uncertainty by reversing the order
of the lepton and $B_{\rm tag}$ selection
and comparing with our standard selection order
using the same MC simulation of our signal model
used to estimate the reconstruction efficiency.
Since we find the reversed selection order efficiency to be compatible
with the standard selection order efficiency within
the precision of our MC simulation, we estimate
the systematic uncertainty as the statistical uncertainty of
that comparison.

\begin{table}[ht]
\centering
\caption{Sources of systematic uncertainties.
}
\label{tab:r24-systematic}
\begin{tabular}{lcc}
\hline\hline
Yield systematics (events) & $\ell = e$ & $\ell = \mu$ \\
\hline
Peaking background: sample statistics
  & $1.0$  & $1.4$ \\
Peaking background: ${\mathcal B}(\Bbar\to\Lambda_c^+X)$ 
  & $1.6$  & $4.7$ \\
Lepton mis-id rate 
  & $0.7$  & $2.0$ \\
Fit bias 
  & $0.3$  & $1.2$ \\
\hline
Total
  & $2.0$  & $5.4$ \\
\hline\hline
Efficiency ratio systematics (\%) & $\ell = e$ & $\ell = \mu$ \\
\hline
Signal model   
  & $11.3$ & $35.9$ \\
Reco. efficiency statistics 
  & $8.4$ & $11.4$ \\
Peaking background: $\Bbar\to\Lambda_c^+X$
  & $1.9$ & $1.9$ \\
Lepton id efficiency 
  & $1.1$  & $2.7$ \\
Selection order 
  & $5.0$ & $6.8$ \\
\hline
Total 
  & $15.1$ & $38.4$ \\
\hline\hline
\end{tabular}
\end{table}

\begin{table}
\caption{Central values of the branching fraction ratio
${{\cal B}(\Bbar \to\Lambda_c^+X\ell^- \overline{\nu}_{\ell})}/
 {{\cal B}(\Bbar \to\Lambda_c^+X)}$.
 The last line averages over $e$ and $\mu$.
}
\label{tab:ratios}
\begin{tabular}{lr@{$\pm$}c@{$\pm$}l}
\hline\hline
Mode & \multicolumn{3}{c}{BF ratio} \\
\hline
$\ell=e$     & $ 2.5$ & $1.1_{\rm stat.}$ & $0.6_{\rm syst.}$ \\
$\ell=\mu$   & $-2.0$ & $2.0_{\rm stat.}$ & $1.9_{\rm syst.}$ \\
$\ell=e,\mu$ & $ 1.7$ & $1.0_{\rm stat.}$ & $0.6_{\rm syst.}$ \\
\hline\hline
\end{tabular}
\end{table}

The central values of the branching fraction ratios are
summarized in Table~\ref{tab:ratios}.
We find a signal significance ${\cal S}=2.1$,
including the systematic uncertainties on the signal yields,
from the difference in the log likelihood values
between the nominal fit and a fit in which we fix the signal yield to zero.
By scanning the likelihood values including the full systematic uncertainties,
we estimate an upper limit at the 90\% confidence:
\begin{equation}
\frac{ {\cal B}(\Bbar \to\Lambda_c^+X\ell^- \overline{\nu}_{\ell})}{{\cal B}(\Bbar \to\Lambda_c^+X)} < 3.5 \%.
\end{equation}

For a comparison with the CLEO result~\cite{Bonvicini:1997fn},
in which the flavor of the semileptonic $B$ was not determined,
we repeat the analysis without requiring the charge-flavor correlation between
the $B_{\rm tag}$ and the $\Lambda_c^+$ in the inclusive mode.
The corresponding yield for the inclusive mode is shown in 
the last row of Table~\ref{tab:correlated}.
We obtain the branching fraction ratio
$
{{\cal B}(\Bbar \to\Lambda_c^+X\ell^- \overline{\nu}_{\ell})}/
     {{\cal B}(B / \Bbar \to\Lambda_c^+X)}=  (1.2\pm 0.7_{\rm stat.}\pm 0.4_{\rm syst.})\%
$
with its corresponding 90\% confidence level upper limit
$
{{\cal B}(\Bbar \to\Lambda_c^+X\ell^- \overline{\nu}_{\ell})}
/{{\cal B}(B / \Bbar \to\Lambda_c^+X)}< 2.5\%,
$
which improves the CLEO limit.
We find that removing the charge-flavor correlation
between the lepton and the $B_{\rm tag}$
in the semileptonic mode also yields consistent results
after re-estimating backgrounds.

In conclusion, we have presented a search for semileptonic $B$ decays into
the charmed baryon $\Lambda_c^+$. We obtain an improved upper 
limit with respect to previous measurements~\cite{Bonvicini:1997fn} on the relative branching fraction
${\cal B}(\Bbar \to\Lambda_c^+X\ell^- \overline{\nu}_{\ell})/
{\cal B}(\Bbar \to\Lambda_c^+X)$, 
which is found to be much smaller than the corresponding
relative branching fraction
for $B$ decays into charmed mesons.
Our result shows that
the rate of baryonic semileptonic $B$ decay is too small to
contribute substantially to the branching fractions of
inclusive semileptonic $B$ decays.

We are grateful for the excellent luminosity and machine conditions
provided by our \pep2\ colleagues, 
and for the substantial dedicated effort from
the computing organizations that support \babar.
The collaborating institutions wish to thank 
SLAC for its support and kind hospitality. 
This work is supported by
DOE
and NSF (USA),
NSERC (Canada),
CEA and
CNRS-IN2P3
(France),
BMBF and DFG
(Germany),
INFN (Italy),
FOM (The Netherlands),
NFR (Norway),
MES (Russia),
MEC (Spain), and
STFC (United Kingdom). 
Individuals have received support from the
Marie Curie EIF (European Union) and
the A.~P.~Sloan Foundation.


\begin{thebibliography}{99}
\bibitem{ell}
Charge conjugation is always implied unless stated otherwise.
\end{thebibliography}
\end{document}